\documentclass[aps,nofootinbib,superscriptaddress,preprint,showpacs]{revtex4}
\usepackage{amsmath}
\usepackage{amssymb}
\usepackage{graphicx}
\begin{document}

\title{Unparticle Physics and Supersymmetry Phenomenology}
\author{Hao Zhang}
\email{haozhang.pku@pku.edu.cn}
\author{Chong Sheng Li}
\email{csli@pku.edu.cn}
\author{Zhao Li}
\email{zhli.phy@pku.edu.cn}

\pacs{11.30.Pb,12.60.-i,12.90.+b,13.90.+i,14.80.-j}

\affiliation{Department of Physics, Peking University, Beijing
100871, China}

 \begin{abstract}
 We show a natural form of the interaction between unparticle and supersymmetry.
 Using the couplings of unparticle to supersymmetry presented,
 as examples, we calculate the differential decay rates for the processes
 ${\it {\tilde f}}\rightarrow {\it f}+{\cal
 U}_{3/2}$, ${\it {\tilde \chi}}^0\rightarrow{\it \gamma}+{\cal
 U}_{3/2}$, ${\it {\tilde \chi}}^\pm\rightarrow W^\pm+{\cal
 U}_{3/2}$ and ${\it {\tilde \chi}}^0\rightarrow Z^0+{\cal
 U}_{3/2}$.
Finally, we discuss the phenomenological implication of our results
and give some comments.
 \end{abstract}
\maketitle
\section{Introduction}
Very recently, Georgi suggested\cite{Georgi:2007ek} that a scale
invariant sector might be treated as an effective theory at TeV
scale. He called it ``unparticle''. Georgi
assumed\cite{Georgi:2007ek} that there is a high energy theory
containing two parts. One contains the fields of the Standard
Model(SM), while the other contains a theory with a nontrivial
infrared fixed point, which will be a scale invariant sector at the
infrared fixed point. An example of such theory is a four
dimensional massless Yang-Mills theory with suitable fermion number.
Actually, more than twenty years ago, Banks and Zaks had
shown\cite{Banks:1981nn} that some massless Yang-Mills theory(${\cal
{BZ}}$ fields) might have a nontrivial infrared fixed point. Georgi
further pointed out\cite{Georgi:2007ek} that if these two parts
interact through the exchange of particles with a large mass scale
${\it M}_{\cal U}$, below the scale ${\it M}_{\cal U}$ a
nonrenormalizable coupling involving both SM fields and ${\cal
{BZ}}$ fields will be induced. And the renormalizable couplings of
the ${\cal {BZ}}$ fields then cause dimensional transmutation as the
scale invariance in the $\cal BZ$ sector emerges at an energy scale
$\Lambda_{\cal U}$, below which the ${\cal BZ}$ operators match onto
unparticle operators. Finally, one can obtain nonrenormalizable
interactions of standard model operators with unparticle operators,
and these effective interactions can be treated in the frame of
effective field theory. Moreover, as shown by
Georgi\cite{Georgi:2007ek}, scale invariance can be used to
calculate the appropriate phase space for unparticle stuff. An
important result is that unparticle stuff with scale dimension ${\it
d}_{\cal U}$ looks like a nonintegral number ${\it d}_{\cal U}$ of
invisible particles\cite{Georgi:2007ek}. To understand this claim,
we remember that the phase space for ${\it n}$ massless particles is
\begin{equation}
(2{\it \pi})^4{\it \delta}^4\left({\it P}-\sum_{{\it j}=1}^{\it
n}{\it {p_j}}\right)\prod_{{\it j}=1}^{\it n}{\it \delta}({\it
{p_j}}^2){\it \theta}({\it {p_j}}^0)\frac{{\it d}^4{\it
{p_j}}}{(2{\it \pi})^3}={\it {A_n}}{\it \theta}({\it P}^0){\it
\theta}({\it P}^2)({\it P}^2)^{{\it n}-2},
\end{equation}
where
\begin{equation}
{\it {A_n}}=\frac{16{\it \pi}^{5/2}}{(2{\it \pi})^{2{\it
n}}}\frac{\Gamma({\it n}+1/2)}{\Gamma({\it n}-1)\Gamma(2{\it n})}.
\end{equation}
For an unparticle operator ${\it O}_{\cal U}$ with dimension ${\it
d}_{\cal U}$, the appropriate phase space can be written
as\cite{Georgi:2007ek}
\begin{equation}
|\langle 0|{\it O}_{\cal U}(0)|{\it P}\rangle|^2{\it \rho}({\it
P}^2)={\it A}_{{\it d}_{\cal U}}{\it \theta}({\it P}^0){\it
\theta}({\it P}^2)({\it P}^2)^{{\it d}_{\cal U}-2},
\end{equation}
with
\begin{equation}
{\it {A_{d_{\cal U}}}}=\frac{16{\it \pi}^{5/2}}{(2{\it \pi})^{2{\it
{d_{\cal U}}}}}\frac{\Gamma({\it {d_{\cal U}}}+1/2)}{\Gamma({\it
{d_{\cal U}}}-1)\Gamma(2{\it {d_{\cal U}}})},
\end{equation}
where $|{\it P}\rangle$ is the unparticle state with 4-momentum $\it
{P^{\mu}}$ produced from the vacuum by ${\it O}_{\cal U}$. This is a
simple result of scale invariance. Using this formula and the
virtual propagator of unparticle\cite{Georgi:2007ek}, one can
calculate some scattering cross sections and decay rates of the
processes with unparticle.

Following Georgi, in a few months, various phenomenological
implications on unparticle physics have been explored by many
groups\cite{Georgi:2007si}\cite{Fox:2007sy}\cite{Greiner:2007hr}.
However, all of these studies are concerned with unparticles which
couple to the SM sector through higher dimensional operators in low
energy effective theory. In this paper we propose a scheme in which
the supersymmetry fields and ${\cal {BZ}}$ fields interact via the
exchange of particles with a large mass scale ${\it M}_{\cal U}$.

In fact, if unparticle appears at TeV scale, it will be worth to
investigate the situation where the visible sector is not only SM
but a TeV new physics, such as supersymmetry or extra dimension. The
Minimal Supersymmetric Standard Model(MSSM) is a popular new physics
candidate at TeV scale although we still do not know what is the TeV
new physics. Only future experiment can answer this question. In
this paper we first try to understand what we will see if the
visible sector in Georgi's scheme\cite{Georgi:2007ek} at TeV scale
is MSSM.

\section{Supersymmetry with unparticle}
As we mentioned above, it is an interesting problem to study a
supersymmetric visible sector in Georgi's scheme. However, there are
many uncertainties to solve this problem. One of them is how MSSM
operator couples to unparticle operator. In fact, we can choose this
coupling freely. Different choices will lead to different
phenomenologies. But we will present the natural couplings here and
show some phenomenological results.

One of useful methods for constructing the couplings between
supersymmetry and unparticle is to couple the supercurrent to an
unparticle operator. As we know, for a Lagrangian density which
includes both chiral and gauge superfields, the general form of the
supercurrent is\cite{Weinberg:2000book}
\begin{eqnarray}
{\it S}^{\it \mu}=&&-\frac{1}{4}\sum_{\it A}{\it f}_{\it
{A\rho\sigma}}[{\it \gamma}^{\it \rho},{\it \gamma}^{\it
\sigma}]{\it \gamma}^{\it \mu}{\it \lambda}_{\it A}-{\it i}\sum_{\it
{Anm}}(\it t_{\it A})_{\it {nm}}{\it \gamma}_\textrm{5}{\it
\gamma}^{\it \mu}{\it \lambda}_{\it A}{{\it \phi}_{\it
n}}^{\ast}{\it \phi}_{\it m}+\frac{\textrm 1}{\sqrt{\textrm
2}}\sum_{\it n}[( \displaystyle{\not}{\it{D}}{\it \phi})_{\it n}{\it
\gamma}^{\it \mu}{\it \psi}_{\it {nR}} \nonumber
\\&&+(\displaystyle{\not}{\it{D}}{\it \phi}^\ast)_{\it
n}{\it \gamma}^{\it \mu}{\it \psi}_{\it {nL}}+{\textrm
2}\left(\frac{{\it {\partial W(\phi)}}}{\it {\partial
\phi_n}}\right){\it \gamma^\mu}{\it \psi_{nL}}+{\textrm
2}\left(\frac{{\it {\partial W(\phi)}}}{\it {\partial
\phi_n}}\right)^\ast{\it \gamma^\mu}{\it \psi_{nR}}],
\end{eqnarray}
where ${\it {f_{\mu\nu}}}$ is gauge field, ${\it \lambda}$ is
gaugino field, ${\it t_A}$ is the generator of the gauge group,
${\it \phi}$ is the scalar component fields of chiral superfields,
the superpartners of ${\it \phi}$ and ${\it \phi^\ast}$ are ${\it
{\psi_L}}$ and ${\it {\psi_R}}$(they are Majorana spinors), ${\it
{W(\phi)}}$ is the superpotential of chiral superfields. Since the
supercurrent is a spin-$\frac{3}{2}$ fermion operator, the
unparticle operator must be a spin-$\frac{3}{2}$ fermion operator
too in order to get a Lorentz invariant interaction. Thus this
interaction can be written as
\begin{equation}
{\it {C_{\cal U}}}{\frac{\it {\Lambda_{\cal U}^{\it {d_{\cal
{BZ}}-d_{\cal U}}}}}{\it {M_{\cal U}^k}}}{\it {{\bar S}_\mu}}{\cal
U}_{3/2}^{\it \mu}+h.c.,
\end{equation}
where ${\cal U}_{3/2}^{\it \mu}$ is the spin-$\frac{3}{2}$
unparticle operator with the scaling dimension ${\it d_{\cal U}}$,
${\it C_{\cal U}}$ is a coefficient function, ${\it d_{\cal BZ}}$ is
mass dimension of $\cal {BZ}$ operator.

We now calculate some sparticles decay processes from Eq.(6). The
general differential decay rate is
\begin{equation}
{\it d}{\Gamma}=\frac{|{\cal M}|^2}{2{\it M}}{\it d}{\Phi}({\it P}),
\end{equation}
where
\begin{equation}
{\it d}{\Phi}({\it P})=\int{(2\pi)^4{\it \delta}^4\left({\it
{P-\sum_j{p_j}}}\right)\prod_{\it j}{{\it d}{\Phi(\it
{p_j})\frac{{\it d}^{\textrm{4}}{\it
{p_j}}}{(\textrm{2}\pi)^\textrm{4}}}}}.
\end{equation}
For an unparticle, the final state density is\cite{Georgi:2007ek}
\begin{equation}
{\it d}{\Phi}_{\cal U}({\it p_{\cal U}})={\it A}_{{\it d}_{\cal
U}}{\it \theta}({\it p_{\cal U}}^0){\it \theta}({\it p_{\cal
U}}^2)({\it p_{\cal U}}^2)^{{\it d}_{\cal U}-2}.
\end{equation}
When we consider the scale invariance violation of the unparticle
stuff, the final state density is\cite{Fox:2007sy}
\begin{equation}
{\it d}{\Phi}_{\cal U}({\it p_{\cal U}})={\it A}_{{\it d}_{\cal
U}}{\it \theta}({\it p_{\cal U}}^0){\it \theta}({\it p_{\cal
U}}^2-\mu^2)({\it p_{\cal U}}^2-\mu^2)^{{\it d}_{\cal U}-2},
\end{equation}
where ${\it \mu}$ is a mass scale at which the scale invariance of
the unparticle stuff is violated. When we calculate the spin sums of
the squared matrix element, we need the polarization sums formula of
the unparticle, which can be written as
\begin{equation}
{\alpha}{\it g^{\mu\nu}}\displaystyle{\not}{\it{p}}+{\beta}({\it
p^{\mu}}{\gamma^\nu}+{\gamma^\mu}{\it p^{\nu}}),
\end{equation}
where ${\it p^\mu}$ is the momentum of the unparticle and $\alpha$
and $\beta$ are free parameters. Using Eqs.(5)-(11), it is
straightforward to calculate several decay rates involving
unparticles in final states as follows.

(1). ${\it {\tilde f}\rightarrow {f+{\cal U}_{\textrm{3/2}}}}$: The
sfermion-fermion-unparticle coupling is given by
\begin{equation}
{\cal {\bar U}}_{3/2\it \mu}(\displaystyle{\not}{\it{\partial}}{\it
{\tilde f}_L^\ast}){\it \gamma}^{\it \mu}{\it {P_L}}{\it f}+{\cal
{\bar U}}_{3/2\it \mu}(\displaystyle{\not}{\it{\partial}}{\it
{\tilde f}_R^\ast}){\it \gamma}^{\it \mu}{\it {P_R}}{\it f}+{\it
{m_f}}{\cal {\bar U}}_{3/2\it \mu}{\it {{\tilde f}_R^\ast}}{\it
\gamma}^{\it \mu}{\it {P_L}}{\it f}+{\it {m_f}}{\cal {\bar
U}}_{3/2\it \mu}{\it {{\tilde f}_L^\ast}}{\it \gamma}^{\it \mu}{\it
{P_R}}{\it f}+h.c..
\end{equation}
The decay width for this process is
\begin{eqnarray}
\Gamma=&&\frac{(4\beta-\alpha){\it A}_{\it d_{\cal U}}{\it C}_{\cal
U}^2{\Lambda}_{\cal U}^{2({\it {d_{\cal BZ}-d_{\cal
U}}})}{m_f}^{d_{\cal U}+1/2}}{2 \pi^2a^{d_{\cal U}+3/2}{\it
{m_{\tilde f}}}^{7/2-d_{\cal U}}{\it M}_{\cal U}^{2{\it k}}}[{\it
a}^2{\it {m_f}}({m_{\tilde f}}^2-{\it {m_f}}^2){\it f}_0({\it
a},{\it d_{\cal U}})\nonumber\\&&+{\it a}{\it {m_{\tilde
f}}}({m_{\tilde f}}^2+{\it {m_f}}^2){\it f}_1({\it a},{\it d_{\cal
U}})-2{m_{\tilde f}}^2{\it {m_f}}{\it f}_2({\it a},{\it d_{\cal
U}})]{\theta}(m_{\tilde f}-m_f-\mu),
\end{eqnarray}
with
\begin{eqnarray}
f_0(a,d)=4^{-d}(1-4 a^2)^{d-1} a^{3/2-d}
   \sqrt{\pi } \Gamma \left(d-\frac{3}{2}\right) \,
   _2\tilde{F}_1\left(\frac{1}{4} (2 d-3),\frac{1}{4}
   (2 d+1);d;1-\frac{1}{4 a^2}\right),
\end{eqnarray}
\begin{eqnarray}
f_1(a,d)=4^{-d}(1-4 a^2)^{d-1} a^{5/2-d}
   \sqrt{\pi } \Gamma \left(d-\frac{3}{2}\right) \,
   _2\tilde{F}_1\left(\frac{1}{4} (2 d-5),\frac{1}{4}
   (2 d+3);d;1-\frac{1}{4 a^2}\right),
\end{eqnarray}
\begin{eqnarray}
f_2(a,d)=-2^{-d-\frac{7}{2}} \sqrt{\pi } \Gamma
   \left(d-\frac{3}{2}\right) G_{3,3}^{2,1}\left(4 a^2|
\begin{array}{c}
 \frac{3}{2},\frac{1}{4} (2 d+5),\frac{1}{4} (2 d+7) \\
 0,2,\frac{5}{2}
\end{array}
\right),
\end{eqnarray}
where ${\it \theta}(z)$ is the Heaviside function, ${\it {m_f}}$ is
the fermion mass, ${\it {m_{\tilde f}}}$ is the sfermion mass,
$a=\frac{{\it {m_f}}{\it {m_{\tilde f}}}}{{\it {m_{\tilde
f}}}^2+{\it {m_f}}^2-\mu^2}$, ${\it {x=\frac{\it {m_{\tilde
f}E_f}}{{\it {m_{\tilde f}}}^\textrm{2}+{\it
{m_f}}^\textrm{2}-\mu^\textrm{2}}}}$ and ${\it {E_f}}$ is the energy
of the final state fermion in the center of mass system.
$_2\tilde{F}_1\left(a,b;c;
  z\right)$ is the regularized hypergeometric function. $G_{p,q}^{m,n}\left(z|
\begin{array}{c}
 a_1,...,a_p \\
 b_1,...,b_q
\end{array}
\right)$ represents the Meijer $G$ function. And the dimensionless
differential decay rate is
\begin{eqnarray}
\frac{{\it {m_{\tilde f}d}}\Gamma}{\Gamma{\it {dE_f}}}=&&a{m_{\tilde
f}}[{\it a}^2{\it {m_f}}({m_{\tilde f}}^2-{\it {m_f}}^2)+{\it a}{\it
{m_{\tilde f}}}({m_{\tilde f}}^2+{\it
{m_f}}^2){\it x}-2{m_{\tilde f}}^2{\it {m_f}}{\it x}^2]\nonumber\\
&&(1-2{\it x})^{{\it d_{\cal U}}-5/2}({\it x}^2-{\it a}^2)^{1/2}{\it
\theta}(\textrm{1}-\textrm{2}{\it x})     {m_f}^{-1}[{\it a}^2{\it
{m_f}}({m_{\tilde f}}^2-{\it {m_f}}^2){\it f}_0({\it a},{\it d_{\cal
U}})\nonumber\\
&&+{\it a}{\it {m_{\tilde f}}}({m_{\tilde f}}^2+{\it {m_f}}^2){\it
f}_1({\it a},{\it d_{\cal U}})-2{m_{\tilde f}}^2{\it {m_f}}{\it
f}_2({\it a},{\it d_{\cal U}})]^{-1}.
\end{eqnarray}

In the limit of the zero fermion mass, the dimensionless
differential decay rate is reduced to
\begin{equation}
\frac{{\it {m_{\tilde f}d}}\Gamma}{\Gamma{\it
{dE_f}}}=\frac{4\Gamma(d+3/2)\Gamma(d+5/2)m_{\tilde f}^2[m_{\tilde
f}^2x^2-2(m_{\tilde f}^2-\mu^2)x^3](1-2 x)^{{\it d}_{\cal
U}-\frac{5}{2}} {\it \theta}(\textrm{1}-\textrm{2}{\it
x})}{\Gamma(d-3/2)(m_{\tilde f}^2-\mu^2)[m_{\tilde
f}^2\Gamma(d+5/2)-3(m_{\tilde f}^2-\mu^2)\Gamma(d+3/2)]}.
\end{equation}

(2). ${\it {\tilde \chi}}^0\rightarrow{\it \gamma}+{\cal
 U}_{3/2}$: The gaugino-photon-unparticle coupling is
\begin{equation}
{\cal {\bar U}}_{3/2\it \mu}{\it {F_{\rho\sigma}^A}}[{\it
{\gamma^\rho}},{\it {\gamma^\sigma}}]{\it {\gamma^\mu}}{\it {\tilde
\chi_A}}+h.c.,
\end{equation}
where $\it {\tilde \chi_A}$ is the gaugino, from which it is easy to
get a mass eigenstate, and $\it {F_{\rho\sigma}^A}$ is the field
strength of the gauge field. The decay width is given by
\begin{eqnarray}
\Gamma=-\frac{3 \alpha {\it A}_{\it d_{\cal U}}{\it C}_{\cal
U}^2{\Lambda}_{\cal U}^{2({\it {d_{\cal BZ}-d_{\cal U}}})}({\it
m_{{\it {\tilde \chi}}^\textrm{0}}}^2-\mu^2)^{{\it d}_{\cal
U}+3/2}\Gamma
   \left({\it
d}_{\cal U}-\frac{3}{2}\right)}{16 \pi^2 {\it m_{{\it {\tilde
\chi}}^\textrm{0}}}^3{\it M}_{\cal U}^{2{\it k}} \Gamma \left({\it
d}_{\cal U}+\frac{5}{2}\right)}{\theta}({\it m_{{\it {\tilde
\chi}}^\textrm{0}}}-\mu),
\end{eqnarray}
where ${\it m_{{\it {\tilde \chi}}^\textrm{0}}}$ is the mass of
neutralino, ${\it x}=\frac{{\it m_{{\it {\tilde
\chi}}^\textrm{0}}}{\it {E_\gamma}}}{{\it m_{{\it {\tilde
\chi}}^\textrm{0}}}^2-{\it \mu}^2}$ and ${\it {E_\gamma}}$ is the
energy of the final state photon in the center of mass system. And
the dimensionless differential decay rate is
\begin{equation}
\frac{{\it m_{{\it {\tilde \chi}}^\textrm{0}}}{\it
d}\Gamma}{\Gamma{\it {dE_\gamma}}}=\frac{8 {\it m_{{\it {\tilde
\chi}}^\textrm{0}}}^2  \Gamma \left({\it d}_{\cal
U}+\frac{5}{2}\right)}{3 \left({\it m_{{\it {\tilde
\chi}}^\textrm{0}}}^2-{\it \mu}^2\right) \Gamma
   \left({\it
d}_{\cal U}-\frac{3}{2}\right)}(1-2 x)^{{\it d}_{\cal
U}-\frac{5}{2}} x^3{\it \theta}(\textrm{1}-\textrm{2}{\it x}).
\end{equation}

(3). ${\it {\tilde \chi}}^0\rightarrow{\it Z^{\textrm{0}}}+{\cal
 U}_{3/2}, {\it {\tilde \chi}}^\pm\rightarrow{\it W^\pm}+{\cal
 U}_{3/2}$: The gaugino-$\it Z^\textrm{0}(\it W^\pm)$-unparticle couplings have been
shown in Eq.(19), and the corresponding decay widths are given by
\begin{eqnarray}
\Gamma=-\frac{\alpha {\it A}_{\it d_{\cal U}}{\it C}_{\cal
U}^2{\Lambda}_{\cal U}^{2({\it {d_{\cal BZ}-d_{\cal
U}}})}{m_G}^{{\it d}_{\cal U}+3/2} }{2 \pi^2 a^{d_{\cal U}+3/2}{\it
m_{\it {\tilde \chi}}}^{5/2-d_{\cal U}}{\it M}_{\cal U}^{2{\it
k}}}[{m_{\tilde \chi}}f_2(a,{\it d}_{\cal U})-a{m_G}f_1(a,{\it
d}_{\cal U})]{\theta}(m_{\it {\tilde \chi}}-m_G-\mu),
\end{eqnarray}
where $m_{\tilde \chi}$ and $m_G$ are the masses of the
neutralino(chargino) and the gauge boson($Z^\textrm{0}$ or $W^\pm$),
respectively, $a=\frac{{\it {m_G}}{\it {m_{\tilde \chi}}}}{{\it
{m_{\tilde \chi}}}^2+{\it {m_G}}^2-\mu^2}$, ${\it {x=\frac{\it
{m_{\tilde \chi}E_G}}{{\it {m_{\tilde \chi}}}^\textrm{2}+{\it
{m_G}}^\textrm{2}-\mu^\textrm{2}}}}$ and ${\it {E_G}}$ is the energy
of the final state of gauge boson in the center of mass system. And
the dimensionless differential decay rates are
\begin{eqnarray}
\frac{{\it {m_{\tilde \chi}d}}\Gamma}{\Gamma{\it
{dE_G}}}=\frac{a{m_{\tilde \chi}}({m_{\tilde
\chi}}x^2-a{m_G}x)}{{m_G}[{m_{\tilde \chi}}f_2(a,{\it d}_{\cal
U})-a{m_G}f_1(a,{\it d}_{\cal U})]}(1-2{\it x})^{{\it d_{\cal
U}}-5/2}({\it x}^2-{\it a}^2)^{1/2}{\it
\theta}(\textrm{1}-\textrm{2}{\it x}).
\end{eqnarray}

\section{Discussions and comments}
\begin{figure}[ht!]
\includegraphics[width=0.6\textwidth]{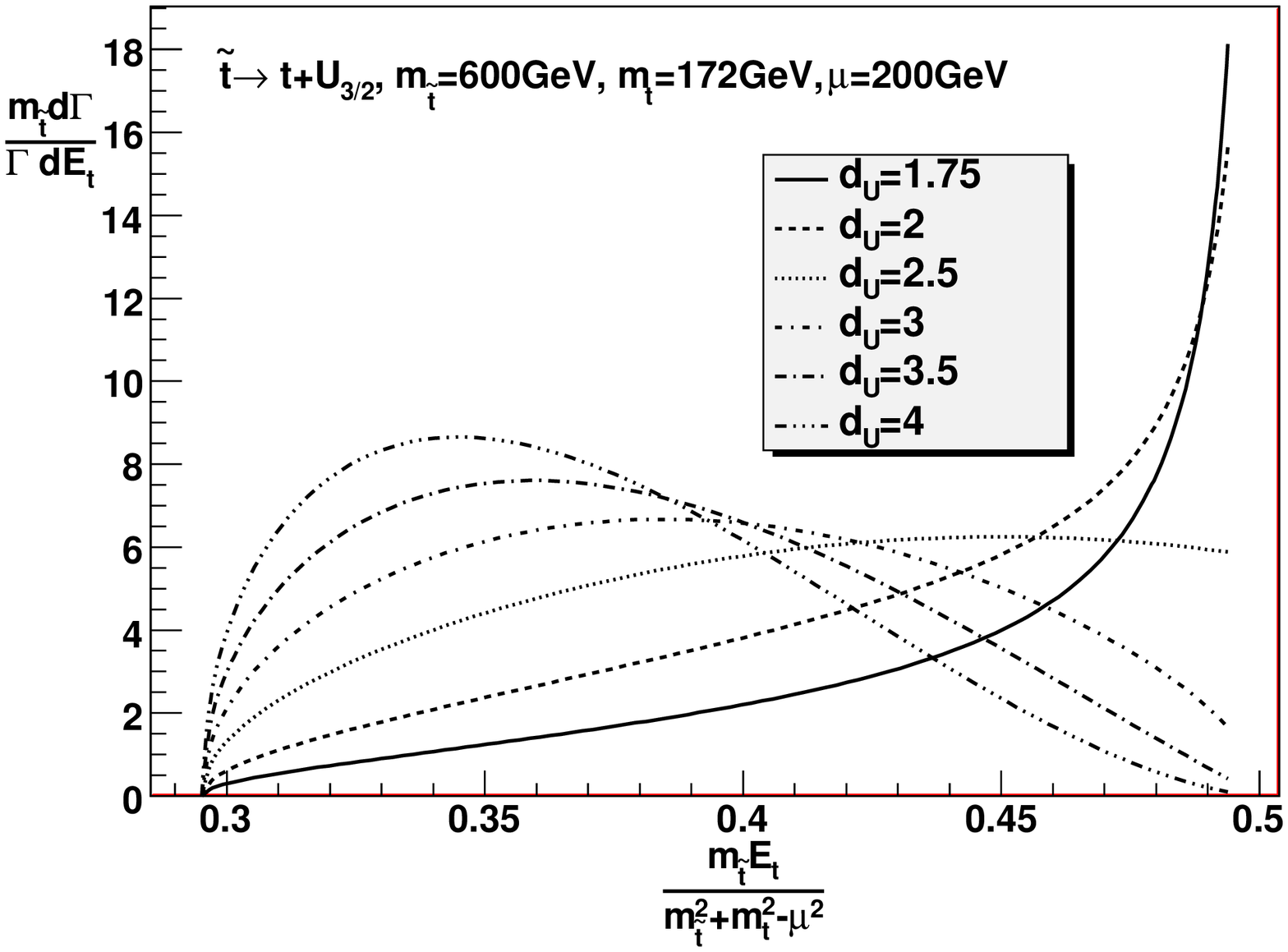}
\caption{The dimensionless differential decay rate $\frac{{\it
{m_{\tilde t}d}}\Gamma}{\Gamma{\it {dE_t}}}$ versus $\frac{{\it
{m_{\tilde t}E_t}}}{{\it {m_{\tilde t}}}^2+{\it {m_t}}^2-\mu^2}$ for
different values of ${{\it d}_{\cal U}}$.} \label{fig1}
\end{figure}

\begin{figure}[ht!]
\includegraphics[width=0.6\textwidth]{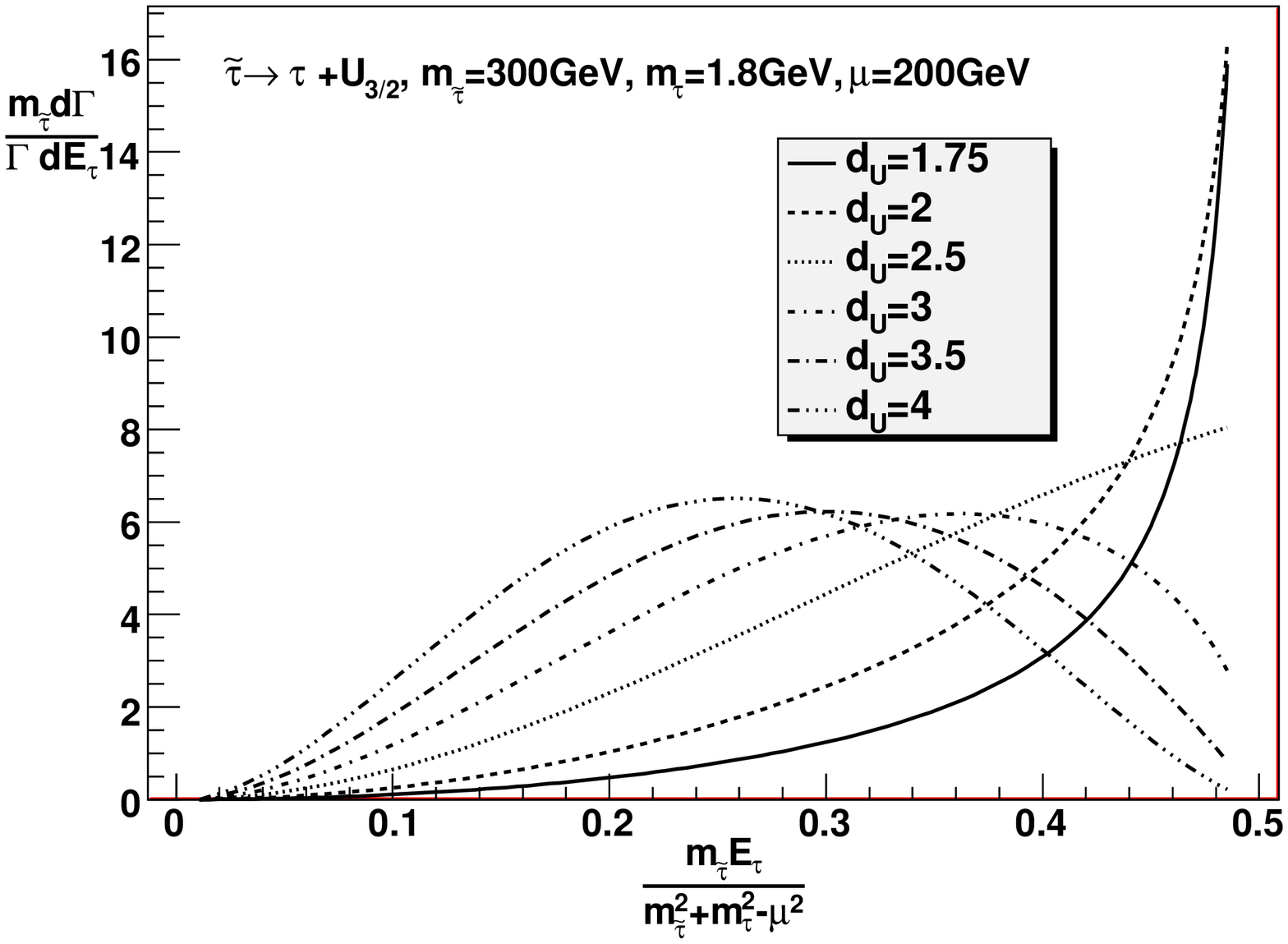}
\caption{The dimensionless differential decay rate $\frac{{\it
{m_{\tilde \tau}d}}\Gamma}{\Gamma{\it {dE_\tau}}}$ versus
$\frac{{\it {m_{\tilde \tau}E_\tau}}}{{\it {m_{\tilde \tau}}}^2+{\it
{m_\tau}}^2-\mu^2}$ for different values of ${{\it d}_{\cal U}}$.}
\label{fig2}
\end{figure}

\begin{figure}[ht!]
\includegraphics[width=0.6\textwidth]{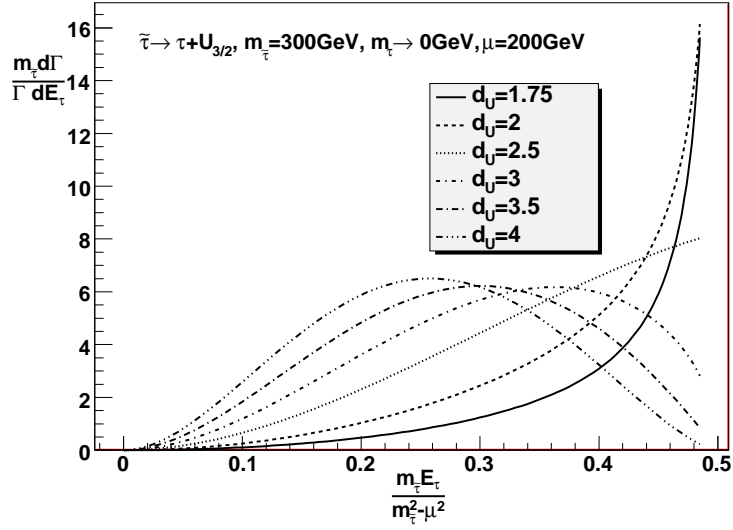}
\caption{The dimensionless differential decay rate $\frac{{\it
{m_{\tilde \tau}d}}\Gamma}{\Gamma{\it {dE_\tau}}}$ versus
$\frac{{m_{\tilde \tau}}{\it {E_t}}}{{\it {m_{\tilde \tau}}}-\mu^2}$
for different values of ${{\it d}_{\cal U}}$ in the zero mass limit
of tau.} \label{fig3}
\end{figure}

We now present some numerical results for the differential decay
rates of sparticles. We first plot the dimensionless differential
decay rate $\frac{{\it {m_{\tilde f}d}}\Gamma}{\Gamma{\it {dE_f}}}$
versus $\frac{{\it {m_{\tilde f}E_f}}}{{\it {m_{\tilde f}}}^2+{\it
{m_f}}^2-\mu^2}$ according to Eqs.(14)-(18) for different masses of
sfermion and fermion, which are shown in Figs.\ref{fig1}-\ref{fig3}.
From these figures, we can see that the shapes of the curves in
Figs.\ref{fig2}-\ref{fig3} are nearly the same, which shows that if
the final state fermion is light, the zero mass limit is a good
approximation.

\begin{figure}[ht!]
\includegraphics[width=0.6\textwidth]{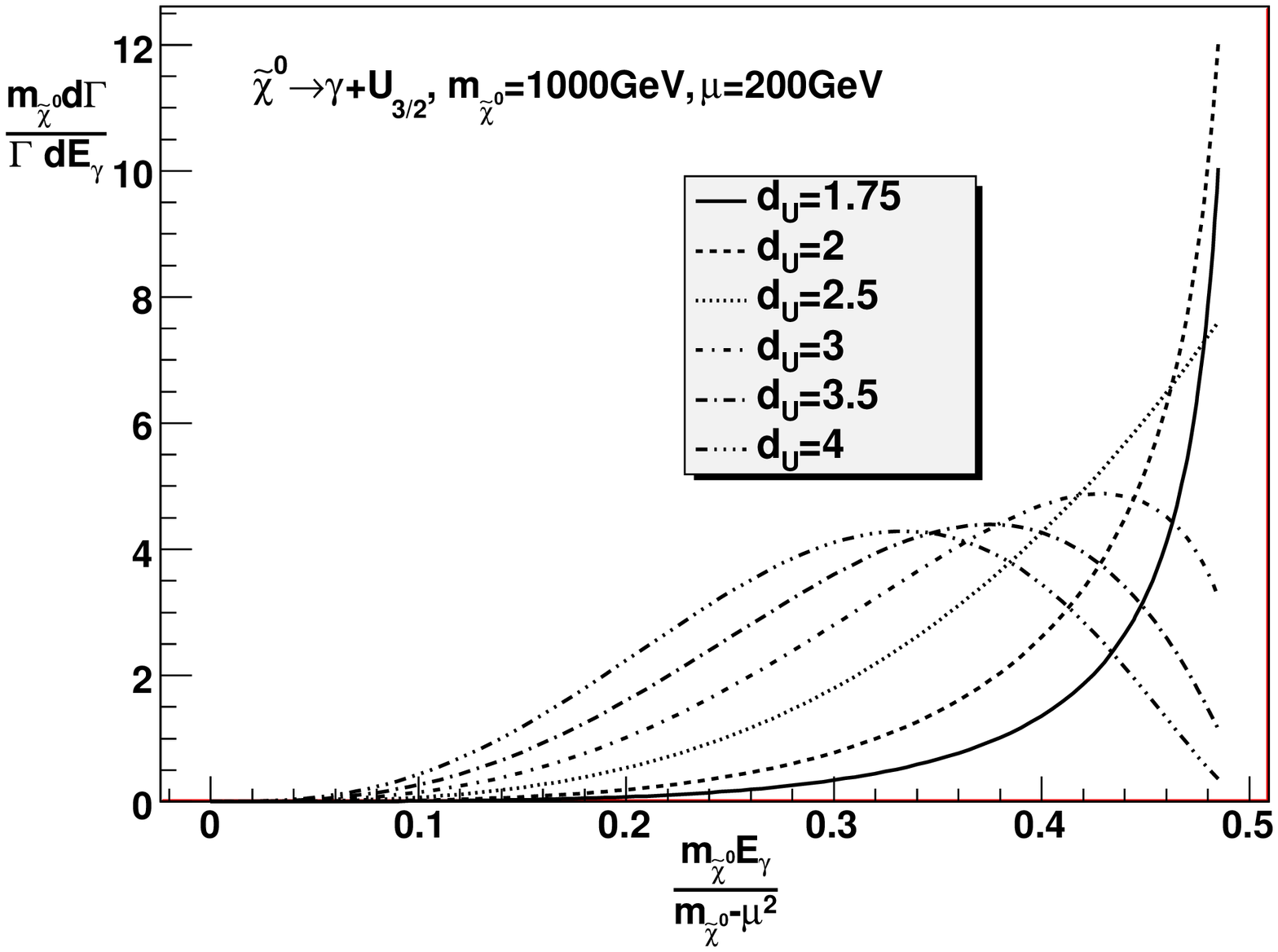}
\caption{The dimensionless differential decay rate $\frac{{ m_{{\it
{\tilde \chi}}^0}}{\it d}\Gamma}{\Gamma{\it {dE_\gamma}}}$ versus
$\frac{m_{{\tilde \chi}^0}E_\gamma}{ m_{{\it {\tilde
\chi}}^0}^2-\mu^2}$ for different values of ${{\it d}_{\cal U}}$.}
\label{fig4}
\end{figure}

It is an interesting signal at future colliders that a neutralino
decays into a single photon plus missing energy. As we have pointed
out, this decay can only happen when the mass of the neutralino is
larger than $\mu$. We calculate the differential decay rate of this
process and show the dependence of the dimensionless differential
decay rate $\frac{{\it m_{{\it {\tilde \chi}}^\textrm{0}}}{\it
d}\Gamma}{\Gamma{\it {dE_\gamma}}}$ on $\frac{ {m_{{\tilde
\chi}^0}E_\gamma}}{\it m_{{\it {\tilde
\chi}}^\textrm{0}}-\mu^\textrm{2}}$ in Fig.\ref{fig4}. Compared with
Fig.\ref{fig3}, it can be found that when ${d_{\cal U}}$ increases
the unparticle carries less energy than one in the sfermion decay
processes.

\begin{figure}[ht!]
\includegraphics[width=0.6\textwidth]{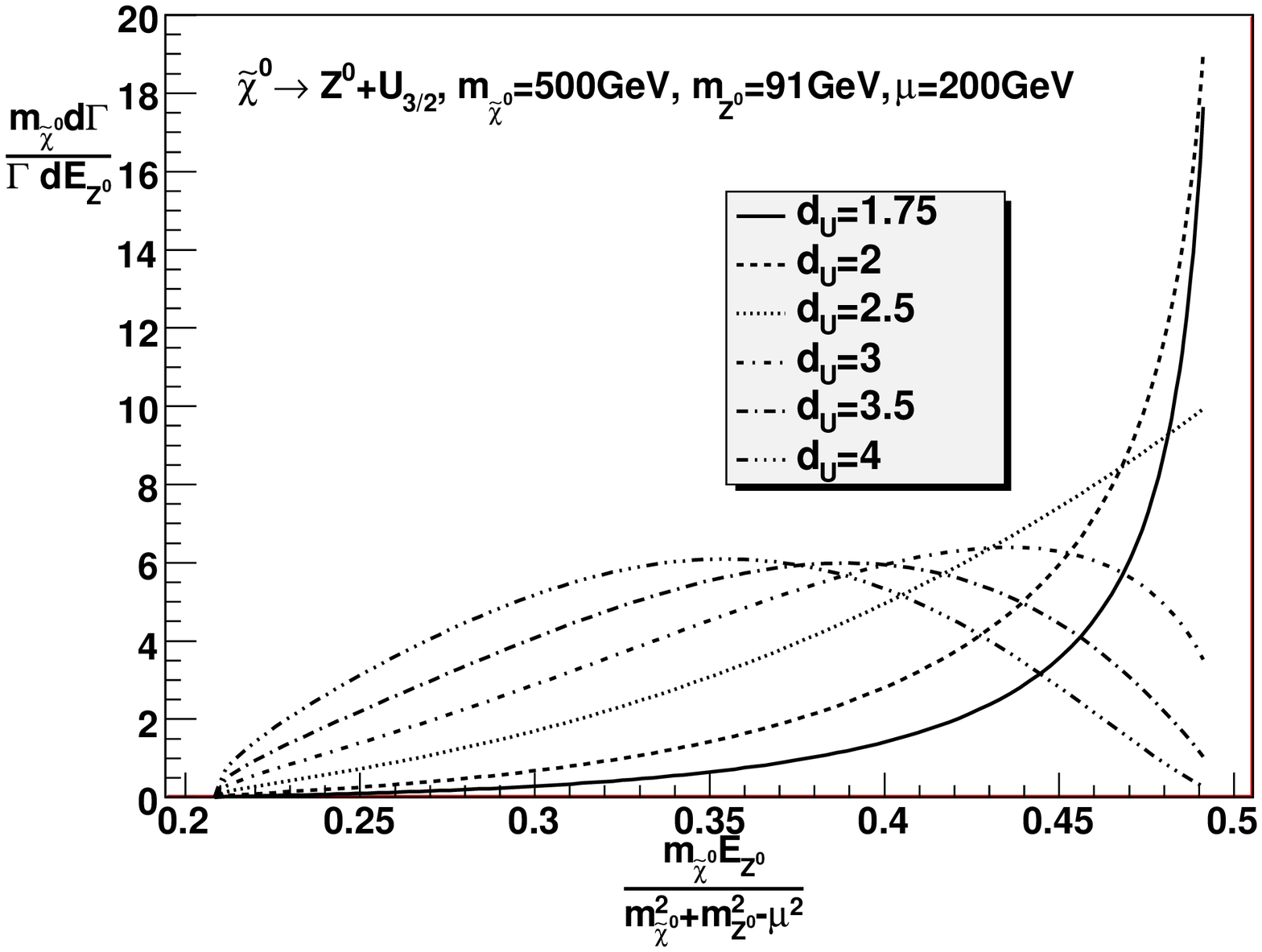}
\caption{The dimensionless differential decay rate $\frac{{
{m_{\tilde \chi^0}d}}\Gamma}{\Gamma{\it {dE_Z}}}$ versus $\frac{{
{m_{\tilde \chi^0}E_Z}}}{{ {m_{\tilde \chi^0}}}^2+{\it
{m_Z}}^2-\mu^2}$ for different values of ${{\it d}_{\cal U}}$.}
\label{fig5}
\end{figure}

\begin{figure}[ht!]
\includegraphics[width=0.6\textwidth]{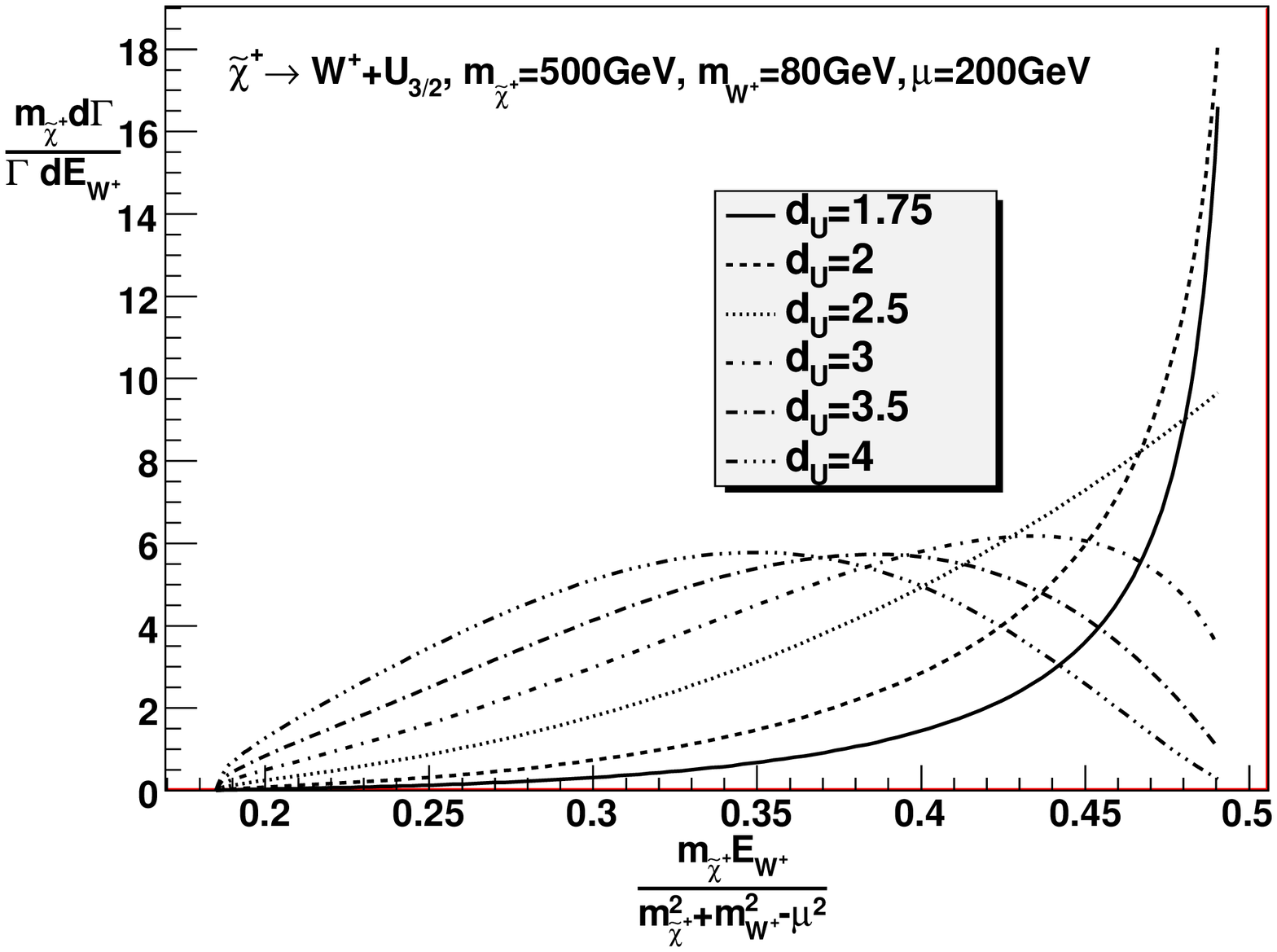}
\caption{The dimensionless differential decay rate $\frac{{\it
{m_{{\tilde \chi}^+}d}}\Gamma}{\Gamma{\it {dE_{W^+}}}}$ versus
$\frac{{\it {m_{{\tilde \chi^+}}E_{W^+}}}}{{\it {m_{\tilde
\chi^+}}}^2+{\it {m_{W^+}}}^2-\mu^2}$ for different values of ${{\it
d}_{\cal U}}$.} \label{fig6}
\end{figure}

The dependence of the dimensionless differential decay rates of
neutralino(chargino) decays into ${Z^0}$(${W^{\pm}}$) plus missing
energy on $\frac{{\it {m_{\tilde \chi}E_G}}}{{\it {m_{\tilde
\chi}}}^2+{\it {m_G}}^2-\mu^2}$ is shown in
Figs.\ref{fig5}-\ref{fig6}. Since the decay can only happen for
${\it {m_{\tilde \chi}}}+{\it {m_G}}\geqslant\mu$, we choose a
relative heavy neutralino(chargino).

In a word, our calculations show that a general property of
sparticle decays into unparticle is that the energy of final state
can change continuously, while the curve of the energy change of
final state should be a delta function in two body decay of a
massive particle in ordinary case. Actually, this is a consequence
of scale invariance. This property could help us detect unparticle
at future colliders. As ${\it d_{\cal U}\rightarrow}\frac{3}{2}$
from above, in general we can see the curves become more peaked at
$\frac{\it {m_{\tilde f}E_f}}{{\it {m_{\tilde f}}}^2+{\it
{m_f}}^2-\mu^2}=0.5$ or $\frac{{\it {m_{\tilde \chi}E_G}}}{{\it
{m_{\tilde \chi}}}^2+{\it {m_G}}^2-\mu^2}=0.5$, as the case shown in
Ref.\cite{Georgi:2007ek}. This means that when the dimension of the
unparticle operator approximates to that of the corresponding
particle operator, the behavior of unparticle looks like one of
particle. We can also find that the shapes of the curves depend
sensitively on $\it d_{\cal U}$, which can help us identify the
dimension of the unparticle operator. If the signals with those
characters are discovered at future colliders, it will be useful for
further understanding both unparticle and supersymmetry.

Finally, we make some comments as following: \vspace{12pt}

1. Although the superpotential of MSSM is still $\it R-$parity
conservation, the $\it R-$parity is apparently violated by the
effective coupling of the supercurrent to the unparticle operator,
which leads that the Lightest Supersymmetry Particle(LSP) in MSSM is
not absolutely stable in this scenario. Thus, the LSP could not be a
good candidate of dark matter. However, if the scale invariance of
unparticle stuff is broken at some energy scale $\mu$, which is
below TeV but larger than the mass of the LSP, the LSP can not decay
into an unparticle, and would be stable. And the LSP could be still
a candidate of dark matter. \vspace{12pt}

2. Since we coupled the unparticle operator to the supercurrent of
MSSM, those couplings break supersymmetry explicitly. Actually,
because the unparticle appears at TeV scale where supersymmetry has
been broken, it dose not take the responsibility for supersymmetry
breaking. Another interesting problem is how the unparticle stuff
affects the invisible sector and the messenger of supersymmetry
breaking\cite{Nakayama:2007nk}. However, this is beyond the scope of
this paper.\vspace{12pt}

3. Obviously, the constraints from electroweak precise observations
to these couplings should be considered. It is also important to
calculate using those couplings some processes at future colliders
and some astrophysics processes which are relevant to dark matter
and ${\it \gamma}$-ray burst. We leave these necessary tasks for a
future work\cite{Zhang:2007pr}. Our interest here was to introduce
new couplings of the unparticle to the supercurrent in MSSM and
discuss some main points concerning its phenomenology.

\begin{acknowledgements}
We would like to thank Wai-Yee Keung for useful discussion. This
work was supported in part by the National Natural Science
Foundation of China, under grants No. 10421503, No. 10575001 and No.
10635030, and the Key Grant Project of Chinese Ministry of Education
under grant No. 305001 and the Specialized Research Fund for the
Doctoral Program of Higher Education.
\end{acknowledgements}
\bibliography{test}

\newpage

\end{document}